\documentstyle[titlepage,12pt]{article}
\begin{document}
\renewcommand{\theequation}{\thesection.\arabic{equation}}
\title{Strings and Multi-Strings in Black Hole and Cosmological
Spacetimes\thanks{Research report to appear in
{\bf "New developments in string gravity and physics at
the Planck energy scale"}.
Edited by N. S\'{a}nchez (World Scientific, Singapore, 1995)}}
\author{A.L. Larsen and N. S\'{a}nchez\\
\\
Observatoire de Paris, DEMIRM.
Laboratoire Associ\'{e} au CNRS\\
UA 336, Observatoire de Paris et
\'{E}cole Normale Sup\'{e}rieure.\\
61, Avenue
de l'Observatoire, 75014 Paris, France.}
\maketitle
\begin{abstract}
Recent results on classical and quantum strings in a
variety of black
hole and cosmological spacetimes, in various dimensions,
are presented. The curved backgrounds under consideration include
the $2+1$ black hole anti de Sitter spacetime and its dual, the black string,
the ordinary $D\geq 4$ black holes with or without a cosmological constant,
the de Sitter and anti de Sitter spacetimes and static Robertson-Walker
spacetimes. Exact solutions to the
string equations of motion and constraints, representing circular strings,
stationary open strings and dynamical straight strings, are obtained
in these backgrounds and
their physical properties (length, energy, pressure) are described.
The existence of
{\it multi-string} solutions, describing finitely or infinitely many strings,
is shown to be a general feature of spacetimes
with a positive or negative cosmological constant.
Generic approximative solutions are obtained using the string perturbation
series approach, and the question of the
stability of the solutions is addressed.

Furthermore, using a canonical quantization procedure, we
find the string mass spectrum in de Sitter and anti de Sitter spacetimes.
New features as compared to the string spectrum in flat Minkowski
spacetime appear, for instance the {\it fine-structure effect} at all levels
beyond the graviton in both de Sitter and anti de Sitter spacetimes,
and the {\it non-existence} of a Hagedorn
temperature in anti de Sitter spacetime.
We discuss the physical implications of these results.
Finally, we consider the effect of spatial curvature on the string
dynamics in Robertson-Walker spacetimes.
\end{abstract}
\section{Introduction}
The classical and quantum propagation of strings in curved spacetimes has
attracted a great deal of interest in recent years. The main complication, as
compared to the case of flat Minkowski spacetime,
is related to the non-linearity
of the equations of motion. It makes it possible to obtain the complete
analytic solution only in a very few special cases like conical spacetime
\cite{san3} and plane-wave/shock-wave backgrounds \cite{san4}. There are
however
also very general results concerning integrability and solvability for
maximally symmetric spacetimes \cite{mic1,four} and gauged WZW models
\cite{bar,six}. These are the exceptional cases; generally the string equations
of motion in curved spacetimes are not integrable and even if they are, it
is usually an extremely difficult task to actually separate the equations,
integrate them and finally write down the complete solution in closed form.
Fortunately, there are several different ways to "attack" a system of coupled
non-linear partial differential equations.

The systematic
study of string dynamics in curved spacetimes and its associated physical
phenomena was
started in Refs.\cite{veg1,san1}. Besides numerical methods, which
will not be discussed here, approximative [7-10]
and exact [11-14]
methods for solving the string equations
of motion and constraints in curved spacetimes,
have been developed. Classical and quantum string dynamics have been
investigated in black hole backgrounds [15-18], cosmological
spacetimes [7, 11-14, 17-21],
cosmic string spacetime \cite{san3}, gravitational
wave backgrounds \cite{san4},
supergravity backgrounds
(which are necessary for fermionic strings) \cite{san5},
and near spacetime singularities \cite{san6}. Physical
phenomena like the Hawking-Unruh
effect in string theory \cite{san1,san7}, horizon
string stretching \cite{san1,san7}, particle
transmutation \cite{san2,san8}, string scattering \cite{san3,san4,san2}, mass
spectrum and critical
dimension \cite{san3,veg1,san2,all2}, string
instability [7, 11-13, 16, 21] and multi-string
solutions [11-13, 28]
have been found.

In a generic $D$-dimensional curved spacetime with metric $g_{\mu\nu}$ and
coordinates $x^\mu,\;(\mu=0,1,...,D-1),$ the string
equations of motion and constraints are:
\begin{equation}
\ddot{x}^\mu-x''^\mu+\Gamma^\mu_{\rho\sigma}(\dot{x}^\rho\dot{x}^\sigma-
x'^\rho x'^\sigma)=0,
\end{equation}
\begin{equation}
g_{\mu\nu}\dot{x}^\mu x'^\nu=g_{\mu\nu}(\dot{x}^\mu\dot{x}^\nu+x'^\mu x'^\nu)
=0,
\end{equation}
where dot and prime stand for derivative with respect to the
world-sheet coordinates $\tau$ and
$\sigma,$ respectively and $\Gamma^\mu_{\rho\sigma}$ are the Christoffel
symbols with respect to the metric $g_{\mu\nu}.$
In the following we present recent results
[26-29] on solutions to Eqs.(1.1)-(1.2) in a variety of
curved spacetimes from cosmology, gravitation and string theory.
We obtain explicit (exact and/or approximate)
mathematical solutions, discuss the corresponding
physical properties, we quantize the solutions in different ways
(canonical quantization, semi-classical quantization) and find the physical
content: the mass spectrum.

The presentation is organized as follows: In Section 2, we discuss the
classical string dynamics in the $2+1$ black hole anti de Sitter (BH-AdS)
spacetime, recently found by Ba\~{n}ados et al \cite{ban1}. We compare with the
string dynamics in ordinary cosmological and black hole spacetimes. This
clarifies the geometry (as seen by a string) of the BH-AdS spacetime.
In Section 3, generalizing results from Section 2, we derive the quantum
string mass spectrum in ordinary $D$-dimensional anti de Sitter
spacetime. We discuss, in particular, the sectors of low and very high mass
states. New physical phenomena arise like the fine-structure effect at all
levels beyond the graviton and the non-existence of a Hagedorn critical
temperature. The results are compared with
corresponding results obtained in Minkowski and de Sitter spacetimes.
Sections 4 and 5 are devoted to the investigation of the more general
underlying structure of solutions to Eqs.(1.1)-(1.2). In Section 4, we
find new classes of exact string and multi-string
solutions in cosmological and black hole
spacetimes, while in Section 5, we consider the effects of a non-zero
spatial curvature on the classical and quantum
string dynamics in Friedmann-Robertson-Walker (FRW)
universes.
\section{Classical String Dynamics in $2+1$ BH-AdS}
\setcounter{equation}{0}
Anti de Sitter (AdS) spacetime is often considered to be a spacetime of
minor importance in a cosmological context. However,
first of all, it {\it is} a FRW universe and as such should not be neglected,
secondly; it serves as a simple and
convenient spacetime for comparison and understanding of results
obtained in (say)
Minkowski or de Sitter spacetimes and third, it has a tendency to
show up (in disguise) as a solution in various models of
dilaton-gravity and string theory. An example of the latter is represented
by the $2+1$ BH-AdS spacetime of Ba\~{n}ados et al \cite{ban1}.
This spacetime background has arised much interest recently
[31-35]. It
describes a two-parameter family (mass $M$ and angular momentum $J$) of black
holes in 2+1 dimensional general relativity with metric:
\begin{equation}
ds^2=(M-\frac{r^2}{l^2})dt^2+(\frac{r^2}{l^2}-M+\frac{J^2}{4r^2})^{-1}dr^2
-Jdtd\phi+r^2 d\phi^2.
\end{equation}
It has two horizons
$r_\pm=\sqrt{\frac{Ml^2}{2}\pm\frac{l}{2}\sqrt{M^2 l^2-J^2}}$ and a
static limit $r_{\mbox{erg}}=\sqrt{M}l,$ defining
an ergosphere, as for ordinary Kerr
black holes. The spacetime is not asymptotically flat; it approaches
anti de Sitter spacetime asymptotically with cosmological constant
$\Lambda=-1/l^2.$ The curvature is constant $R_{\mu\nu}=-(2/l^2)g_{\mu\nu}$
everywhere, except probably at $r=0,$ where it has at most a delta-function
singularity. Notice the weak nature of the singularity at $r=0$ in 2+1
dimensions as compared with the power law divergence of curvature scalars in
$D>3$ (We will not discuss here the geometry near $r=0.$ For a discusion,
see Refs.\cite{ban2,delta}). The
spacetime, Eq.(2.1), is also a solution of the low
energy effective action of string theory with zero dilaton field $\Phi=0,$
anti-symmetric tensor field $H_{\mu\nu\rho}=(2/l^2)\epsilon_{\mu\nu\rho}\;\;
(\mbox{i.e.}\;B_{\phi t}=r^2/l^2)$ and $k=l^2$ \cite{hor1}. Moreover, it
yields an
exact solution of string theory in 2+1 dimensions, obtained by gauging the
WZWN sigma model of the group $SL(2,R)\times R$ at level
$k$ \cite{hor1,kal} (for
non-compact groups, $k$ does not need to be an integer, so the central charge
$c=3k/(k-2)=26$ when k=52/23). This solution is the black string background
\cite{hor2}:
\begin{eqnarray}
&d\tilde{s}^2=-(1-\frac{{\cal M}}{\tilde{r}})d\tilde{t}^2+(1-\frac{{\cal Q}^2}
{{{\cal M}}\tilde{r}})d\tilde{x}^2+(1-\frac{{\cal M}}{\tilde{r}})^{-1}
(1-\frac{{\cal Q}^2}{{{\cal M}}\tilde{r}})^{-1}\frac{l^2 d\tilde{r}^2}
{4\tilde{r}^2},&\nonumber
\end{eqnarray}
\begin{eqnarray}
&\tilde{B}_{\tilde{x}\tilde{t}}=\frac{{\cal Q}}{r},\;\;\;
\tilde{\Phi}=-\frac{1}{2}
\log\tilde{r}l,&
\end{eqnarray}
which is related by duality \cite{hor1,bus} to
the 2+1 BH-AdS spacetime, Eq.(2.1). It
has two horizons $\tilde{r}_{\pm}=r_\pm,$ the same as the metric, Eq.(2.1),
while the static limit is $\tilde{r}_{\mbox{erg}}=J/(2\sqrt{M}).$

We first investigate the string propagation in the $2+1$ BH-AdS background
by considering
the perturbation series around the exact center of mass of the string:
\begin{equation}
x^\mu(\tau,\sigma)=q^\mu(\tau)+\eta^\mu(\tau,\sigma)+\xi^\mu(\tau,\sigma)+...
\end{equation}
The original method of Refs.\cite{veg1,san1} can be conveniently formulated in
covariant form. It is useful to introduce $D-1$ normal vectors
$n^\mu_R\;\;(R=1,..,D-1),$ (which can be chosen to be covariantly constant
by gauge fixing), and consider comoving perturbations $\delta x_R,$ i.e. those
seen by an observer travelling with the center of mass, thus $\eta^\mu=
\delta x^Rn^\mu_R.$ After Fourier expansion, $\delta x^R(\tau,\sigma)=
\sum_n C^R_n(\tau)e^{-in\sigma},$ the first order perturbations satisfy the
matrix Schr\"{o}dinger-type equation in $\tau$:
\begin{equation}
\ddot{C}_{nR}+(n^2\delta_{RS}-R_{\mu\rho\sigma\nu}n^\mu_R n^\nu_S
\dot{q}^\rho\dot{q}^\sigma)C^S_n=0.
\end{equation}
Second order perturbations $\xi^\mu$ and constraints are similarly covariantly
treated, $\xi^\mu$ also satisfying Schr\"{o}dinger-type equations
with source terms, see Ref.\cite{all1}.

For our purposes here it is enough to consider the non-rotating $(J=0)$ 2+1
BH-AdS background and a radially infalling string. We have solved completely
the
c.m. motion $q^\mu(\tau)$ and the first and second order perturbations
$\eta^\mu(\tau,\sigma)$ and $\xi^\mu(\tau,\sigma)$ in this background.
Eqs.(2.4) become:
\begin{equation}
\ddot{C}_{nR}+(n^2+\frac{m^2}{l^2})C_{nR}=0;\;\;\;R=\perp,\parallel
\end{equation}
The first order perturbations are independent of the black hole mass, only
the anti de Sitter (AdS) part emerges. All oscillation frequencies $\omega_n=
\sqrt{n^2+m^2/l^2}$ are real; there are no unstable modes in this case. The
perturbations:
\begin{equation}
\delta x_R(\tau,\sigma)=\sum_n
[ A_{nR}e^{-i(n\sigma+\omega_n\tau)}+\tilde{A}_{nR}
e^{-i(n\sigma-\omega_n\tau)}]
\end{equation}
are completely finite and regular. This is also true for the second order
perturbations which are bounded everywhere, even for $r\rightarrow 0\;(\tau
\rightarrow 0).$ We have also computed the conformal generators $L_n,$ (see
Ref.\cite{all1}),
and the string mass:
\begin{equation}
m^2=2\sum_n (2n^2+\frac{m^2}{l^2})[A_{n\parallel}
\tilde{A}_{-n\parallel}+
A_{n\perp}\tilde{A}_{-n\perp}].
\end{equation}
The mass formula is modified (by the term $m^2/l^2$) with respect to the
usual flat spacetime expression. This is due to the asymptotic (here AdS)
behaviour of the spacetime. In ordinary $D\geq 4$ black hole spacetimes
(without
cosmological constant), which are asymptotically flat, the mass spectrum
is the same as in flat Minkowski spacetime \cite{san2}.

We compare with the string perturbations in the ordinary ($D\geq 4$) black hole
anti de Sitter spacetime. In this case Eqs.(2.4) become:
\begin{equation}
\ddot{C}_{nS\perp}+(n^2+m^2 H^2+\frac{Mm^2}{r^3})
C_{nS\perp}=0,\;\;\;S=1,2
\end{equation}
\begin{equation}
\ddot{C}_{n\parallel}+(n^2+m^2 H^2-\frac{2Mm^2}{r^3})
C_{n\parallel}=0.
\end{equation}
The transverse $\perp$-perturbations are oscillating with real frequencies
and are bounded even for $r\rightarrow 0.$ For
longitudinal $\parallel$-perturbations, however, imaginary
frequencies arise and instabilities develop.
The $(\mid n\mid=1)$-instability sets in at:
\begin{equation}
r_{\mbox{inst.}}=(\frac{2Mm^2}{1+m^2H^2})^{1/3}
\end{equation}
Lower modes become unstable even outside the horizon, while higher modes
develop instabilities at smaller $r$ and eventually only for $r\approx 0.$
For $r\rightarrow 0$ (which implies $\tau\rightarrow\tau_0$) we find $r(\tau)
\approx (3m\sqrt{M/2})^{2/3}(\tau_0-\tau)^{2/3}$ and:
\begin{equation}
\ddot{C}_{nS\perp}+\frac{2}{9(\tau-\tau_0)^2}C_{nS\perp}=0,\;\;\;S=1,2
\end{equation}
\begin{equation}
\ddot{C}_{n\parallel}-\frac{4}{9(\tau-\tau_0)^2}C_{n\parallel}=0.
\end{equation}
For $\tau\rightarrow\tau_0$ the $\parallel$-perturbations blow up while
the string
ends trapped into the $r=0$ singularity. We see the important difference
between the string evolution in the 2+1 BH-AdS background and the ordinary
3+1 (or higher dimensional) black hole anti de Sitter spacetime.

We also compare with the string propagation in the 2+1 black string background,
Eq.(2.2) (with $J=0$). In this case, Eqs.(2.4) become:
\begin{equation}
\ddot{C}_{n\perp}+n^2 C_{n\perp}=0,
\end{equation}
\begin{equation}
\ddot{C}_{n\parallel}+(n^2-\frac{2m^2M}{lr})C_{n\parallel}=0.
\end{equation}
The $\perp$-modes are stable, while $C_{n\parallel}$ develop instabilities. For
$r\rightarrow 0$ (which implies $\tau\rightarrow\tau_0$) we find $r(\tau)
\approx\frac{m^2M}{l}(\tau_0-\tau)^2$ and:
\begin{equation}
\ddot{C}_{n\parallel}-\frac{2}{(\tau_0-\tau)^2}C_{n\parallel}=0,
\end{equation}
with similar conclusions as for the ordinary 3+1 (or higher dimensional)
black hole anti
de Sitter spacetime.
\vskip 6pt
\hspace*{-6mm}In order to extract more information about the string evolution
in these backgrounds, in particular exact properties, we consider the
circular string ansatz:
\begin{equation}
t=t(\tau),\;\;\;r=r(\tau),\;\;\;\phi=\sigma+f(\tau),
\end{equation}
in the equatorial plane $(\theta=\pi/2)$ of the stationary axially symmetric
backgrounds:
\begin{equation}
ds^2=g_{tt}(r)dt^2+g_{rr}(r)dr^2+2g_{t\phi}(r)dtd\phi+g_{\phi\phi}(r)d\phi^2.
\end{equation}
This includes
all the cases of interest here: The 2+1 BH-AdS spacetime, the black
string, as well as the equatorial plane of
ordinary Einstein black holes. The string dynamics, determined by
Eqs.(1.1)-(1.2), is
then reduced to a system of second order ordinary differential equations and
constraints, also described as a Hamiltonian system:
\begin{equation}
\dot{r}^2+V(r)=0;\;\;\;\;\;V(r)=g^{rr}(g_{\phi\phi}+E^2g^{tt}),
\end{equation}
\begin{equation}
\dot{t}=-Eg^{tt},\;\;\;\;\;\dot{f}=-Eg^{t\phi};\;\;\;\;E=-P_t=\mbox{const.},
\end{equation}
which in all backgrounds considered
here are solved in terms of either elementary
or elliptic functions. The dynamics of the circular strings takes place at
the $r$-axis in the $(r,V(r))$-diagram and from the properties of the potential
$V(r)$ (minima, zeroes, asymptotic behaviour for large $r$ and the value
$V(0)$), general knowledge about the string motion can be
obtained. On the other hand, the line element of the circular string turns
out to be:
\begin{equation}
ds^2=g_{\phi\phi}(d\sigma^2-d\tau^2),\;\;\;\mbox{i.e.}\;\;S(\tau)=\sqrt
{g_{\phi\phi}(r(\tau))},
\end{equation}
$S(\tau)$ being the invariant string size. For all the static black hole AdS
spacetimes ( 2+1 and higher dimensional): $\;S(\tau)=r(\tau).$

For the rotating 2+1 BH-AdS spacetime:
\begin{equation}
V(r)=r^2(\frac{r^2}{l^2}-M)+\frac{J^2}{4}-E^2,
\end{equation}
(see Fig.1). $V(r)$ has a global minimum $V_{\mbox{min}}<0$ between the two
horizons $r_+,r_-$ (for $Ml^2\geq J^2,$ otherwise there are no horizons). The
vanishing of $V(r)$ at $r=r_{01,2}$ (see Ref.\cite{all1}) determines three
different types of solutions: (i) For $J^2>4E^2,$ there are two
positive zeroes $r_{01}<r_{02},$ the string never comes outside the static
limit, never falls into $r=0$ neither (there is a barrier between $r=r_{01}$
and $r=0$). The mathematical solution oscillates between $r_{01}$ and
$r_{02}$ with $0<r_{01}<r_-<r_+<r_{02}<r_{\mbox{erg}}.$ It may be
interpreted as a string travelling between the different universes described
by the maximal analytic extension of the manifold. (ii) For $J^2<4E^2,$ there
is only one positive zero $r_0$ outside the static limit and there is no
barrier preventing the string from collapsing into $r=0$. The string starts
at $\tau=0$ with maximal size $S^{(ii)}_{\mbox{max}}$ outside the static
limit, it then contracts through the ergosphere and the two horizons
and eventually collapses into a point $r=0.$ For $J\neq 0,$ it may be
still possible to continue this solution into another universe like in
the case (i). (iii) $J^2=
4E^2$ is the limiting case where the maximal string size equals the static
limit: $S^{(iii)}_{\mbox{max}}=l\sqrt{M}.$ In this case $V(0)=0,$ thus the
string contracts through the two horizons and eventually collapses into a
point $r=0.$

The exact general solution in the three cases (i)-(iii) is given by:
\begin{equation}
r(\tau)=\mid r_m-\frac{1}{c_1\wp(\tau-\tau_0)+c_2}\mid,
\end{equation}
where:
\begin{equation}
r_m=S_{\mbox{max}}=\sqrt{\frac{Ml^2}{2}}\sqrt{1+\sqrt{1-\frac{4V(0)}{Ml^2}}},
\;\;\;V(0)=\frac{J^2}{4}-E^2.
\end{equation}
$c_1,c_2$ are constants in terms of $(l,M,r_m),$ given in Ref.\cite{all1}, and
$\wp$ is the Weierstrass elliptic $\wp$-function with invariants $(g_2,g_3),$
discriminant $\Delta$ and roots $(e_1,e_2,e_3),$ also given in Ref.\cite{all1}.
The three cases (i)-(iii) correspond to the cases $\Delta>0,$ $\Delta<0$ and
$\Delta=0,$ respectively. Notice that
$S^{(ii)}_{\mbox{max}}>S^{(iii)}_{\mbox{max}}=l\sqrt{M}>S^{(i)}_{\mbox{max}}.$
In the case (i), $r(\tau)$ can be written in terms of the Jacobian elliptic
function $\mbox{sn}[\tau^*,k],\;\tau^*=\sqrt{e_1-e_3}\;\tau,\;k=\sqrt{(e_2-
e_3)/(e_1-e_3)}.$ It
follows that the solution (i) oscillates between the two
zeroes $r_{01}$ and $r_{02}$ of $V(r),$ with
period $2\omega,$ where $\omega$ is
the real semi-period of the Weierstrass
function: $\omega=K(k)/\sqrt{e_1-e_3}\;,$ in
terms of the complete elliptic integral of the first kind $K(k).$ We have:
\begin{equation}
r(0)=r_m,\;\;\;r(\omega)=\sqrt{Ml^2-r_m^2},\;\;\;r(2\omega)=r_m,...
\end{equation}
In the case (ii) ($\Delta<0$) two roots ($e_1,e_3$) become complex, the string
collapses into a point $r=0$ and we have:
\begin{equation}
r(0)=r_m,\;\;\;r(\frac{\omega_2}{2})=0,\;\;\;r(\omega_2)=r_m,...
\end{equation}
where $\omega_2$ is the real semi-period of the Weierstrass function for this
case. In the case (iii) ($\Delta=0$) the elliptic functions reduce to
hyperbolic functions:
\begin{equation}
r(\tau)=\frac{\sqrt{M}l}{\cosh(\sqrt{M}\tau)},
\end{equation}
so that:
\begin{equation}
r(-\infty)=0,\;\;\;r(0)=r_m=\sqrt{M}l,\;\;\;r(+\infty)=0.
\end{equation}
Here, the string starts as a point, grows until $r=r_m$
(at $\tau=0$), and then it contracts until it collapses again ($r=0$) at
$\tau=+\infty.$ In this case the string makes only
one oscillation between $r=0$ and $r=r_m.$

Notice that for the static background ($J=0$), the only allowed motion is (ii),
i.e. $r_m>r_{\mbox{hor}}=\sqrt{M}l$ (there is no ergosphere and only one
horizon in this case), with:
\begin{equation}
r_m=\sqrt{\frac{Ml^2}{2}}\sqrt{1+\sqrt{1+\frac{4E^2}{Ml^2}}}.
\end{equation}
For $J=0,$ the string collapses into $r=0$ and stops there. The Penrose
diagram of the 2+1 BH-AdS spacetime for $J=0$ is very similar to the
Penrose diagram of the ordinary ($D\geq 4$) Schwarzschild  spacetime, so the
string motion outwards from $r=0$ is unphysical because of the causal
structure. The coordinate time $t(\tau)$ can be expressed in terms of the
incomplete elliptic integral of the third kind $\Pi,$ see Ref.\cite{all1}. The
string has its maximal size $r_m$ at $\tau=0,$ passes the horizon at
$\tau=\tau_{\mbox{hor}}$ (expressed in terms of an incomplete elliptic
integral of the first kind) and falls into $r=0$ for $\tau=
\omega_2/2,$ $\omega_2$ being the real semi-period of the Weierstrass
function. That is, we have:
\begin{eqnarray}
&r(0)=r_m,\;\;\;r(\tau_{\mbox{hor}})=\sqrt{M}l,\;\;\;r(\omega_2/2)
=0,&\nonumber\\
&t(0)=0,\;\;\;t(\tau_{\mbox{hor}})=\infty,&
\end{eqnarray}
and $t(\omega_2/2)$ can be expressed in terms of the Jacobian zeta function
$Z,$ see Ref.\cite{all1}.
\vskip 6pt
\hspace*{-6mm}We also study the circular strings in the ordinary $D\geq 4$
spacetimes. In the generic
3+1 Kerr anti de Sitter (or Kerr de Sitter) spacetime,
the potential $V(r)$ is however quite complicated, see Ref.\cite{all1}.
The general circular string solution involves higher genus elliptic functions
and it is not necessary to go into details here. We will compare with the
non-rotating cases, only.

It is instructive to recall \cite{vil4} the circular
string in Minkowski spacetime (Min), for
which $V(r)=r^2-E^2$ (Fig.2a), the string
oscillates between its maximal size $r_m=E,$ and $r=0$ with the solution
$r(\tau)=r_m\mid\cos\tau\mid.$

In the Schwarzschild black hole (S)
$V(r)=r^2-2Mr-E^2$ (Fig.2b), the solution is remarkably simple: $r(\tau)=M+
\sqrt{M^2+E^2}\cos\tau.$ The mathematical solution oscillates between
$r_m=M+\sqrt{M^2+E^2}$ and $M-\sqrt{M^2+E^2}<0,$ but because of the causal
structure and the curvature singularity the motion can not be continued after
the string has collapsed into $r=0.$

For anti de Sitter spacetime (AdS), we
find: $V(r)=r^2(1+H^2r^2)-E^2$ (Fig.2c). The
string oscillates between $r_m=\frac{1}{\sqrt{2}H}\sqrt{-1+\sqrt{1+4H^2E^2}}$
and $r=0$ with the solution:
\begin{equation}
r(\tau)=r_m\mid\mbox{cn}[(1+4H^2E^2)^{1/4},k]\mid,
\end{equation}
which is periodic with period $2\omega:$
\begin{equation}
\omega=\frac{K(k)}{(1+4H^2E^2)^{1/4}},\;\;\;\;k=\sqrt{\frac{\sqrt{1+4H^2E^2}-1}
{2\sqrt{1+4H^2E^2}}}.
\end{equation}

For Schwarzschild anti de Sitter spacetime (S-AdS), we find $V(r)=
r^2(1+H^2r^2)-2Mr-E^2$ (Fig.2d) and:
\begin{equation}
r(\tau)=r_m-\frac{1}{d_1\wp(\tau)+d_2},\;\;\;\;r(0)=r_m
\end{equation}
$d_1,d_2$ are constants given in terms of $(M,H,r_m),$ see Ref.\cite{all1}
($r_m$ is the root of the equation $V(r)=0$,  which has
in this case exactly one positive solution). The invariants, the
discriminant and the roots are also given explicitly in Ref.\cite{all1}.
The string starts with $r=r_m$
at $\tau=0,$ it then contracts and eventually collapses into the $r=0$
singularity. The existence of elliptic function solutions for the
string motion is characteristic of the presence of a cosmological constant.
For $\Lambda=0=-3H^2$ the circular string motion reduces to simple
trigonometric functions. From Fig.2 and our analysis we see that the circular
string motion is
qualitatively very similar in all these backgrounds (Min, S, AdS,
S-AdS): the string has a maximal {\it bounded} size
and then it contracts towards
$r=0.$ There are however physical and quantitative differences: in Minkowski
and pure anti de Sitter spacetimes, the string truly oscillates between
$r_m$ and $r=0,$ while in the black hole cases (S, S-AdS), there is only one
half oscillation, the string motion stops at $r=0.$ This also holds for the
2+1 BH-AdS spacetime with $J=0$ (Fig.1b). Notice also that in all these
cases, $V(0)=-E^2<0$ and $V(r)\sim r^\alpha;\;(\alpha=2,4)$ for $r>>E.$

The similarity can be pushed one step further by considering small
perturbations
around the circular strings. We find \cite{all}:
\begin{equation}
\ddot{C}_{n}+(n^2+\frac{r}{2}\frac{da(r)}{dr}+\frac{r^2}{2}\frac{d^2a(r)}
{dr^2}-\frac{2E^2}{r^2})C_n=0,
\end{equation}
determining the Fourier components of the comoving perturbations. For the
spacetimes of interest here, $a(r)=1-2M/r+H^2r^2$ (Min, S, AdS, S-AdS), or
$a(r)=r^2/l^2-1$ (2+1 BH-AdS), the comoving perturbations are regular except
near $r=0,$ where we find (for all cases) $r(\tau)\approx-E(\tau-\tau_0)$ and:
\begin{equation}
\ddot{C}_{n}-\frac{2}{(\tau-\tau_0)^2}C_n=0.
\end{equation}
It follows that not only the unperturbed circular strings, but
also the comoving
perturbations around them behave in a similar way in all these non-rotating
backgrounds (2+1 and higher dimensional). This should be contrasted with the
string perturbations around the center of mass, which behave differently in
these backgrounds. It must be noticed that for rotating ($J\neq 0$) spacetimes,
the circular string behaviour is qualitatively different from the non-rotating
($J=0$) spacetimes. For large $J,$ both in the 2+1 BH-AdS as well as in the
3+1 ordinary Kerr-AdS spacetimes, non-collapsing circular string solutions
exist. The potential $V(r)\rightarrow +\infty$ for $r\rightarrow 0$ and no
collapse into $r=0$ is possible.

The dynamics of circular strings in curved spacetimes is determined by the
string tension, which tends to contract the string, and by the local gravity
(which may be attractive or repulsive). In all the previous non-rotating
backgrounds, the
local gravity is attractive (i.e. $da(r)/dr>0$), and it acts together with
the string tension in the sense of contraction. But in spacetimes with regions
in which repulsion (i.e. $da(r)/dr<0$) dominates, the strings can expand with
unbounded radius (unstable strings \cite{mic,veg4}). It
may also happen that the
string tension and the local gravity be of the same order, i.e. the
two opposite
effects can balance, and the string is stationary. De Sitter spacetime provides
an example in which all such type of solutions exist \cite{mic,veg4}. In
de Sitter
spacetime, $V(r)$ is unbounded from below for $r\rightarrow\infty$ ($V(r)\sim
-r^4$) and unbounded expanding circular strings are present. In addition, an
interesting new feature appears in the presence of a
positive cosmological constant: the existence of {\it multi-string}
solutions [9-11]. The
world-sheet time $\tau$ turns out to be a multi
(finite or infinite) valued function of the physical time. That is, one single
world-sheet where $-\infty\leq\tau\leq+\infty,$ can describe many (even
infinitely many \cite{veg4})
different and independent strings (in flat spacetime,
one single world-sheet describes only one string). In the S, AdS and S-AdS
spacetimes, the multi-string feature for circular strings is absent.

The main conclusions of this section are given in Tables 1, 2. Further
details can be found in Ref.\cite{all1}
\section{Quantum String Spectrum in Ordinary AdS}
\setcounter{equation}{0}
In the first part of the
previous section, the classical string dynamics was studied in
the $2+1$ BH-AdS
spacetime. In this section we go further in the investigation of the physical
properties of strings in AdS spacetime, by performing string quantization.
Since the $2+1$ BH-AdS spacetime is locally AdS,
the results of Section 2
can be extended to classical strings
in $D$-dimensional AdS spacetime as well. In AdS
spacetime, the string motion is oscillatory in time and is {\it stable}; all
fluctuations around the string center of mass are well behaved and bounded.
Local
gravity of AdS spacetime
is always negative and string instabilities do not develop.
The string perturbation series approach, considering fluctuations around the
center of mass, is particularly appropriate in AdS spacetime,
the natural
dimensionless expansion parameter being $\lambda=\alpha'/l^2>0,$ where
$\alpha'$
is the string tension and the `Hubble constant' $H=1/l.$ The
negative cosmological
constant of AdS spacetime is related to the `Hubble constant' $H$ by
$\Lambda=-(D-1)(D-2)H^2/2.$

All the spatial ($\mu=1,...,D-1$) modes in $D$-dimensional AdS,
oscillate with frequency
$\omega_n=\sqrt{n^2+m^2\alpha'\lambda},$ which are
real for all $n$ ($m$ being the string mass). In this section, we
perform a canonical
quantization procedure. From the conformal generators $L_n,\;\tilde{L}_n$ and
the constraints $L_0=\tilde{L}_0=0,$ imposed at the quantum level, we obtain
the mass formula (the quantum version of Eq.(2.7) written in units
where $\alpha'$ appears explicitly):
\begin{equation}
m^2\alpha'=(D-1)\sum_{n>0}\Omega_n(\lambda)
+\sum_{n>0}\Omega_n(\lambda)\sum_{R=1}^{D-1}[(\alpha^R_n)^{\dag}
\alpha^R_n+(\tilde{\alpha}^R_n)^{\dag}\tilde{\alpha}^R_n],
\end{equation}
where:
\begin{equation}
\Omega_n(\lambda)=\frac{2n^2+m^2\alpha'\lambda}{\sqrt{n^2+m^2\alpha'\lambda}},
\end{equation}
and we have applied symmetric ordering of the operators. The operators
$\alpha^R_n,\;\tilde{\alpha}^R_n$ satisfy:
\begin{equation}
[\alpha^R_n,\;(\alpha^R_n)^{\dag}]=[\tilde{\alpha}^R_n,\;(\tilde{\alpha}^R_n)^
{\dag}]=1,\;\;\;\;\;\;\mbox{for all}\;\;n>0.
\end{equation}
For $\lambda<<1,$ which is clearly
fulfilled in most interesting cases, we have found the lower mass states
$m^2\alpha'\lambda<<1$ and the quantum mass spectrum \cite{all2}.
Physical states are
characterized by the eigenvalue of the number operator:
\begin{equation}
N=\frac{1}{2}\sum_{n>0} n\sum_{R=1}^{D-1}[(\alpha^R_n)^{\dag}
\alpha^R_n+(\tilde{\alpha}^R_n)^{\dag}\tilde{\alpha}^R_n],
\end{equation}
and the ground state is defined by:
\begin{equation}
\alpha^R_n\mid 0>=\tilde{\alpha}^R_n\mid 0>=0,\;\;\;\;\;\;
\mbox{for all}\;\;n>0.
\end{equation}
We find that $m^2\alpha'=0$ is an {\it exact} solution
of the mass formula in $D=25$ and that there is a graviton at $D=25,$ which
indicates, as in de Sitter space \cite{veg1}, that the critical dimension
of AdS is 25 (although it should be stressed that the question whether these
spacetimes are solutions to the full $\beta$-function equations remains open).
As in Minkowski spacetime,
the ground state is a tachyon.
Remarkably enough, for $N\geq 2$ we find that a generic feature of all excited
states beyond the graviton, is the presence of a {\it fine structure} effect:
for a given eigenvalue $N\geq 2,$ the corresponding states have different
masses. For the lower mass states the expectation value of the mass operator in
the corresponding
states (generically labelled $\mid {j}>$) turns out to have the
form (see Ref.\cite{all2}):
\begin{equation}
<{j}\mid m^2\alpha' \mid {j}>_{\mbox{AdS}}=a_{{j}}
+b_{{j}}\lambda^2+c_{{j}}\lambda^3+{\cal O}(\lambda^4).
\end{equation}
The collective index "j" generically labels the state $\mid {j}>$ and the
coefficients $a_{{j}},\;b_{{j}},\;c_{{j}}$ are all well computed numbers,
different for each state, their precise values are given in Ref.\cite{all2}.
The corrections to the mass in Minkowski spacetime
appear to order $\lambda^2.$ Therefore, the leading Regge trajectory for
the lower mass states is:
\begin{equation}
J=2+\frac{1}{2}m^2\alpha'+{\cal O}(\lambda^2).
\end{equation}
In Minkowski spacetime the mass
and number operator of the string are related by
$m^2\alpha'=-4+4N.$ In AdS (as well as in de Sitter (dS)), there is no such
simple relation between the mass and the number operators; the splitting of
levels increases considerably for very large $N.$ The {\it fine structure}
effect we find here is also present in de Sitter space.
Up to order $\lambda^2,$ the lower mass states in dS
and in AdS are the same, the differences appear to the order $\lambda^3.$ The
lower mass states in de Sitter spacetime are given by Eq.(3.6) but with the
$\lambda^3$-term getting an opposite sign ($-c_{{j}}\lambda^3$).

For the very high mass spectrum, we find more drastic effects. States with
very large eigenvalue $N,$ namely $N>>1/\lambda,$ have masses:
\begin{equation}
<j\mid m^2\alpha' \mid j>_{\mbox{AdS}}\approx d_{{j}}\lambda N^2
\end{equation}
and angular momentum:
\begin{equation}
J^2\approx\frac{1}{\lambda}m^2\alpha',
\end{equation}
where $d_{{j}}$ are well computed numbers different for each state,
their precise values are given in Ref.\cite{all2}.
Since $\lambda=\alpha'/l^2,$ we see from Eq.(3.8) that the masses of the high
mass states are {\it independent} of $\alpha'.$
In Minkowski spacetime, very large $N$ states all have the same mass
$m^2\alpha'\approx 4N,$ but here in AdS the masses of the
high mass states with the same
eigenvalue $N$ are {\it different} by factors $d_{{j}}.$
In addition, because of the fine structure effect, states with different $N$
can get mixed up. For high mass states, the level spacing {\it grows} with $N$
(instead of being constant as in Minkowski spacetime). As a consequence, the
density of states $\rho(m)$ as a function of mass grows like
$\mbox{Exp}[(m/\sqrt{\mid\Lambda\mid}\;)^{1/2}]$ (instead of
$\mbox{Exp}[m\sqrt{\alpha'}]$ as in Minkowski spacetime), and
{\it independently}
of $\alpha'.$ The partition function for a gas of strings at a temperature
$\beta^{-1}$ in AdS spacetime
is well defined for all finite temperatures $\beta^{-1},$
discarding the existence of the
Hagedorn temperature.

For the low mass states ($m^2\alpha'\lambda<<1$) in anti de Sitter spacetime,
our results can be written as \cite{all2}:
\begin{equation}
<j\mid m^2(\alpha',l)\mid j>=\frac{4(N-1)}{\alpha'}+\frac{1}{l^2}
\sum_{n=0}^{\infty}a_{jn}(\alpha'/l^2)^n,
\end{equation}
where "$j$" is a collective index labelling the state $\mid j>.$ It is now
important to notice that $a_{j0}=0$ for {\it all}
the low mass states  \cite{all2}, i.e.
there is no "constant" term on the right hand side of Eq.(3.10). A non-zero
$a_{j0}$-term would give rise to a $\alpha'$-independent contribution to
the string mass. Its absense, on the other hand, means that the first term
on the right hand side of Eq.(3.10) is super-dominant (since, in all cases,
$\alpha'/l^2=\lambda<<1$) and that the string scale is therefore set by
$1/\alpha'.$

For the high mass states
($m^2\alpha'\lambda>>1$) we found instead  \cite{all2}:
\begin{equation}
<j\mid m^2(\alpha',l)\mid j>\approx\frac{d_j}{l^2}N^2,\;\;\;\;\;\;\mbox{for}\;
N>>l^2/\alpha'
\end{equation}
where the number $d_j$ depends on the state.
The right hand side of Eq.(3.11) is exactly like a non-zero dominant
$a_{j0}$-term in Eq.(3.10). For the high mass states the scale is therefore
set by $1/l^2$ which is equal to the absolute value of the cosmological
constant $\Lambda$ (up to a geometrical factor) and
{\it independent} of $\alpha'.$ This suggests that for
$\lambda<<1,$ the masses of {\it all} string states can be represented by a
formula of the form (3.10). For the low mass states $a_{j0}=0,$ while for
the high mass states $a_{j0}$ becomes a large positive number.

In the black string background, Eq.(2.2),
we have calculated explicitly the first and second
order string fluctuations around the center of mass \cite{all2}.
We then determined the
world-sheet energy-momentum tensor and we derived the mass formula in the
asymptotic region. The mass spectrum is equal to the mass spectrum in flat
Minkowski spacetime. Therefore, for a gas of strings at temperature
$\beta^{-1}$ in the asymptotic region of the black string background,
the partition function goes like:
\begin{equation}
Z(\beta)\sim\int^{\infty} dm\; e^{-m(\beta-\sqrt{\alpha'}\;)},
\end{equation}
which is only defined for $\beta>\sqrt{\alpha'},$ i.e. there is a Hagedorn
temperature:
\begin{equation}
T_{\mbox{Hg}}=(\alpha')^{-1/2}.
\end{equation}
In higher dimensional ($D\geq 4$) black hole spacetimes the next step now
would be to set up a scattering formalism, where a string from an asymptotic
in-state interacts with the gravitational field of the black hole and reappears
in an asymptotic out-state \cite{san2}. However, this is not possible in the
black string background. In the uncharged black string
background under consideration here, {\it all} null and timelike geodesics
incoming from spatial infinity pass through the horizon and fall into the
physical singularity \cite{hor2}. No "angular momentum", as in the case of
scattering off the ordinary Schwarzschild black hole, can
prevent a point particle from falling into the singularity. The string
solutions considered here are based on perturbations around
the string center of mass which follows, at least approximately,
a point particle geodesic. Therefore, a string incoming from
spatial infinity
inevitably falls into the singularity in the black string background.
\section{New Classes of Exact String and Multi-String
Solutions in Curved Spacetimes}
\setcounter{equation}{0}
In this section we return to the string equations of motion and constraints,
Eqs.(1.1)-(1.2), to look for new classes of exact solutions.
In most spacetimes, quite general families of
exact solutions can be found by making an appropriate ansatz, which exploits
the symmetries of the underlying curved spacetime. In axially symmetric
spacetimes, a convenient ansatz corresponds to circular strings, as
we saw in Section 2. Such an
ansatz effectively decouples the dependence on the spatial world-sheet
coordinate $\sigma,$ and the string equations of motion and constraints
reduce to non-linear coupled ordinary differential equations,
Eqs.(2.18)-(2.19). In
this section we will make instead
an ansatz which effectively decouples the dependence on the temporal
world-sheet coordinate $\tau.$ This ansatz, which we call the "stationary
string ansatz" is dual to the "circular string ansatz" in the sense that it
corresponds to a formal interchange of the world-sheet coordinates
$(\tau,\sigma),$ as well as of the azimuthal angle $\phi$ and the
stationary time $t$ in the target space:
\begin{equation}
\tau\;\leftrightarrow\;\sigma,\;\;\;\;\;\;\;\;t\;\leftrightarrow\;\phi.
\end{equation}
The stationary string ansatz will describe stationary strings when $t$
(and $\tau$) are timelike, for instance in anti de Sitter spacetime
(in static coordinates) and outside the horizon of a Schwarzschild black hole.
On the other hand, if $t$ (and $\tau$) are spacelike, for instance inside the
horizon of a Schwarzschild black hole or outside the horizon of de Sitter
spacetime (in static coordinates), the stationary string ansatz will describe
dynamical propagating strings. Considering for simplicity
a static line element in the form:
\begin{equation}
ds^2=-a(r)dt^2+\frac{dr^2}{a(r)}+r^2(d\theta^2+\sin^2\theta d\phi^2),
\end{equation}
the stationary string ansatz reads explicitly:
\begin{equation}
t=\tau,\;\;\;\;\;r=r(\sigma),\;\;\;\;\;\phi=\phi(\sigma),\;\;\;\;\;\theta=
\pi/2.
\end{equation}
The string equations of motion and constraints, Eqs.(1.1)-(1.2),
reduce to two separated
first order ordinary differential equations:
\begin{equation}
\phi'=\frac{L}{r^2},\;\;\;\;\;\;r'^2+V(r)=0;\;\;\;\;
V(r)=-a(r)[a(r)-\frac{L^2}{r^2}],
\end{equation}
where $L$ is an integration constant.
The qualitative features of the possible string configurations can be read off
directly from the shape of the potential $V(r).$ Thereafter, the detailed
analysis of the quantitative features can be performed by explicitly solving
the (integrable) system of equations (4.4). The induced line element on the
world-sheet is given by:
\begin{equation}
ds^2=a(r)(-d\tau^2+d\sigma^2).
\end{equation}
Thus, if $a(r)$ is negative, the world-sheet coordinate $\tau$ becomes
spacelike while $\sigma$ becomes timelike and the stationary string
ansatz (4.3) describes a dynamical string. If $a(r)$ is positive, the ansatz
describes a stationary string.

In
this section we solve explicitly Eqs.(4.4) and we analyze in detail the
solutions and their physical interpretation in Minkowski, de Sitter,
anti de Sitter, Schwarzschild and $2+1$ black hole anti de Sitter spacetimes.
In all these cases, the solutions are expressed in terms of elliptic
(or elementary) functions. We furthermore analyze the physical properties,
energy and pressure, of these solutions.
The energy and pressure densities of the strings can be obtained from the $3+1$
dimensional spacetime energy-momentum tensor:
\begin{equation}
\sqrt{-g}T^{\mu\nu}=\frac{1}{2\pi\alpha'}\int d\tau d\sigma
(\dot{X}^\mu\dot{X}^\nu-X'^\mu X'^\nu)\delta^{(4)}(X-X(\tau,\sigma)).
\end{equation}
After integration over a spatial volume that completely encloses the string
\cite{energy},
the energy-momentum tensor takes the form of a
fluid:
\begin{equation}
T^\mu\;_\nu=\mbox{diag.}(-E,P_1,P_2,P_3).
\end{equation}

In Minkowski spacetime (Min), the potential defined in Eqs.(4.4) is given by:
\begin{equation}
V_{\mbox{M}}(r)=\frac{L^2}{r^2}-1,
\end{equation}
and the solution of Eqs.(4.4) describes one infinitely long straight string
with "impact-parameter" $L.$ The equation of state, relating energy and
pressure densities, takes the well-known form Ref.\cite{all3}:
\begin{equation}
dE=-dP_2=\mbox{const.},\;\;P_1=P_3=0.
\end{equation}
In anti de Sitter spacetime (AdS), the potential is given by:
\begin{equation}
V_{\mbox{AdS}}(r)=-(1+H^2r^2)[1+H^2r^2-\frac{L^2}{r^2}].
\end{equation}
The radial coordinate $r(\sigma)$ is periodic with finite period $T_\sigma,$
which can be expressed in terms of a complete elliptic integral,
see Ref.\cite{all3}. For
$\sigma\in[0,T_\sigma],$ the solution describes an infinitely long stationary
string in the wedge $\phi\in\;]0,\Delta\phi[\;,$ where:
\begin{equation}
\Delta\phi=2k\sqrt{\frac{1-2k^2}{1-k^2}}[\Pi(1-k^2,\;k)-K(k)]\;\in\;\;]0,\;\pi[
\end{equation}
The elliptic modulus $k,$ which is a function of $HL$ \cite{all3},
parametrizes the solutions, $k\in\;]0,1/\sqrt{2}[\;.$
The azimuthal angle is generally not a periodic function of $\sigma,$
thus when the spacelike world-sheet coordinate $\sigma$ runs through the range
$]-\infty,+\infty[\;,$ the solution describes an {\it infinite number}
of infinitely long stationary open strings. The general solution is therefore
a {\it multi-string} solution. Until now multi-string solutions were only found
in de Sitter spacetime [11-13]. Our
results show that multi-string
solutions are a general feature of spacetimes with a cosmological
constant (positive or negative). The solution in anti de Sitter spacetime
describes a {\it finite} number of strings if the following relation holds:
\begin{equation}
N\Delta\phi=2\pi M.
\end{equation}
Here $N$ and $M$ are integers, determining the number of strings and the
winding in azimuthal angle, respectively, for the multi-string solution,
see Fig.3. The
equation of state for a full multi-string solution takes the form $(P_3=0)$:
\begin{equation}
dP_1=dP_2=-\frac{1}{2}dE,\;\;\;\;\;\mbox{for}\;\;r\rightarrow\infty
\end{equation}
corresponding to extremely unstable strings \cite{ven}.

In de Sitter spacetime (dS), the potential is given by:
\begin{equation}
V_{\mbox{dS}}(r)=-(1-H^2r^2)[1-H^2r^2-\frac{L^2}{r^2}].
\end{equation}
In this case we have to distinguish between solutions inside the horizon (where
$\tau$ is timelike) and solutions outside the horizon (where $\tau$ is
spacelike). Inside the horizon, the generic
solution describes one infinitely long open
stationary string winding around $r=0.$
For special values of the constants of motion, corresponding to a relation,
which formally takes the same form as Eq.(4.12), the solution describes a
closed string of finite length $l=N\pi/H.$ The integer $N$ in this case
determines the number of "leaves", see Fig.4. The energy is positive and
finite and grows with $N.$ The pressure turns out to vanish identically,
thus the equation of state corresponds to cold matter.

Outside the horizon, the world-sheet coordinate $\tau$ becomes spacelike
while $\sigma$ becomes timelike, thus we define:
\begin{equation}
\tilde{\tau}\equiv\sigma,\;\;\;\;\;\;\;\;\tilde{\sigma}\equiv\tau,
\end{equation}
and the string solution is conveniently
expressed in hyperboloid coordinates.
The radial coordinate $r(\tilde{\tau})$ is periodic with a finite period
$T_{\tilde{\tau}}.$ For $\tilde{\tau}\in[0,T_{\tilde{\tau}}],$
the solution describes a straight string incoming non-radially
from spatial infinity, scattering at the horizon and escaping towards infinity
again, Fig.5. The string length is zero at the horizon and grows
indefinetely in the asymptotic regions.
As in the case of anti de Sitter spacetime,
the azimuthal angle is generally not a periodic function,
thus when the timelike
world-sheet coordinate $\tilde{\tau}$ runs through the range
$]-\infty,+\infty[\;,$ the solution describes an {\it infinite} number
of dynamical straight strings scattering at the horizon at different
angles. The general solution is therefore
a multi-string solution. In particular, a multi-string solution describing a
{\it finite} number of strings is obtained if a relation of the form (4.12) is
fulfilled. It turns out that the solution describes {\it at least}
three strings.
The energy and pressures of a full multi-string solution
have also been computed. In the
asymptotic region they fulfill an equation of state corresponding to extremely
unstable strings, i.e. like Eq.(4.13).

In the Schwarzschild black hole background (S), the potential is given by:
\begin{equation}
V_{\mbox{S}}(r)=-(1-2m/r)[1-2m/r-\frac{L^2}{r^2}].
\end{equation}
No multi-string
solutions are found in this case.
Outside the horizon the solution of Eqs.(4.4) describes one infinitely long
stationary open string. This solution was already
derived in Ref.\cite{fro1}, and we shall not discuss it here.
Inside the horizon, where $\tau$ becomes spacelike
while $\sigma$ becomes timelike, we make the redefinitions (4.15) and the
solution is conveniently expressed in terms of
Kruskal coordinates, see Ref.\cite{all3}. The
solution describes one straight string infalling {\it non-radially} towards the
singularity, see Fig.6. At the horizon, the string length is zero and it grows
{\it indefinitely} when the string approaches the spacetime
singularity.  Thereafter,
the solution can not be continued, see Ref.\cite{all3}.

In the $2+1$ black hole anti de Sitter spacetime (BH-AdS) \cite{ban1},
the potential is given by:
\begin{equation}
V_{\mbox{BH-AdS}}(r)=-(\frac{r^2}{l^2}-1)[\frac{r^2}{l^2}-1-\frac{L^2}{r^2}].
\end{equation}
Outside the horizon, the solutions "interpolate" between the solutions
found in anti de Sitter spacetime and outside the horizon of the
Schwarzschild black hole. The solutions thus describe infinitely long
stationary open strings. As in anti de Sitter spacetime, the general solution
is a {\it multi-string} describing {\it infinitely}
many strings. In particular, for certain
values of the constants of motion, corresponding to the condition of the form
(4.12), the solution describes a finite number of strings. In the simplest
version of the $2+1$ BH-AdS background ($M=1,\;J=0$),
it turns out that the solution describes {\it at least} seven
strings, see Ref.\cite{all3}. Inside the
horizon, we make the redefinitions (4.15) and the solution is conveniently
expressed in
terms of Kruskal-like coordinates, see Ref.\cite{all3}. The solution is
similar to the solution found inside the horizon of the
Schwarzschild black hole, but there is one important difference at $r=0.$
As in the Schwarzschild black hole background, the solution
describes one straight string infalling non-radially towards $r=0,$ and
beyond  this
point the solution can not be continued because of the global structure of the
spacetime. At the horizon the string size is zero and during the fall
towards $r=0,$ the string size {\it grows}
but stays {\it finite}. This should be
compared with the straight string inside the horizon of the
Schwarzschild black hole, where the string size grows indefinetely. The
physical reason for this difference is that the point $r=0$ is not a
strong curvature singularity in the $2+1$ BH-AdS spacetime.

We have considered also the Schwarzschild de Sitter and
Schwarzschild anti de Sitter spacetimes. These spacetimes contain all
the features of the spacetimes already discussed: singularities, horizons,
positive or negative cosmological constants. All the various types of
string solutions found in the other spacetimes (open, closed, straight,
finitely and infinitely long, multi-strings) are therefore present in
the different regions of the Schwarzschild-de Sitter and
Schwarzschild-anti de Sitter spacetimes. The details are given in
Ref.\cite{all3}.

We close this section
with a few remarks on the stability of the solutions. Generally, the
question of stability must be addressed by considering small perturbations
around the exact solutions. In Ref.\cite{fro}, a covariant formalism describing
physical perturbations propagating along an arbitrary string configuration
embedded in an arbitrary curved spacetime, was developed. The resulting
equations determining the evolution of the perturbations are however very
complicated in the general case, although partial (analytical) results
have been obtained in special cases for de Sitter
\cite{all,all6} and Schwarzschild black hole \cite{fro,all6,all7}
spacetimes. The exact solutions found in this section fall
essentially into two classes: dynamical and stationary. The dynamical string
solutions outside the horizon of de Sitter (or S-dS) and inside the horizon
of Schwarzschild (or S-dS, S-AdS) spacetimes, are already unstable at the
zeroth order approximation (i.e. without including small perturbations), in
the sense that their physical length grows indefinetely. For the stationary
string solutions the situation is more delicate. The existence of the
stationary configurations is based on an exact balance between the
string tension and the local attractive or repulsive gravity. For that reason,
it can be expected that the configurations are actually unstable for certain
modes of perturbation, especially in strong curvature regions.

The main conclusions of this section are given in Table 4. Further
details can be found in Ref.\cite{all3}.
\section{Spatial Curvature Effects on the String Dynamics}
\setcounter{equation}{0}
The propagation of strings in Friedmann-Robertson-Walker (FRW)
cosmologies has been
investigated using both exact and approximative methods, see for example
Refs.[7, 11-14, 17-20]
(as well as numerical methods, which shall not be discussed here).
Except for anti de Sitter spacetime, which has negative spatial curvature, the
cosmologies that have been considered until now, have all
been spatially flat. In
this section we will consider the physical effects of a non-zero (positive or
negative) curvature index on the classical and quantum strings. The
non-vanishing components of the Riemann tensor for the generic D-dimensional
FRW line element, in comoving coordinates:
\begin{equation}
ds^2=-dt^2+a^2(t)\frac{d\vec{x} d\vec{x}}{(1+\frac{K}{4}\vec{x}\vec{x})^2},
\end{equation}
are given by:
\begin{equation}
R_{itit}=\frac{-aa_{tt}}{(1+\frac{K}{4}\vec{x}\vec{x})^2},\;\;\;\;\;\;
R_{ijij}=\frac{a^2(K+a_{t}^2)}{(1+\frac{K}{4}\vec{x}\vec{x})^4};\;\;\;\;i\neq j
\end{equation}
where $a=a(t)$ is the scale factor and $K$ is the curvature index. Clearly, a
non-zero curvature index introduces a non-zero spacetime curvature; the
exceptional case provided by $K=-a_{t}^2=\mbox{const.},$ corresponds to
the Milne-Universe. From Eqs.(5.2), it is also seen that the curvature index
has to compete with the first
derivative of the scale factor. The effects of the
curvature index are therefore most conveniently discussed in the family of
FRW-universes with constant scale factor, the so-called static
Robertson-Walker spacetimes. This is the point of view we take in the
present section.

We consider both the closed ($K>0$) and the hyperbolic ($K<0$) static
Robertson-Walker spacetimes, and all our results are compared with the
already known results in the flat ($K=0$) Minkowski spacetime. We determine the
evolution of circular strings, derive the corresponding equations of state
(using Eq.(4.6)),
discuss the question of strings as self-consistent solutions to the
Einstein equations \cite{self}, and we perform a semi-classical
quantization. We also find all the stationary string configurations in these
spacetimes.
\vskip 6pt
\hspace*{-6mm}The radius of a classical circular string in the spacetime
(5.1), for $a=1,$ is determined by:
\begin{equation}
\dot{r}^2+V(r)=0;\;\;\;\;\;\;\;\;V(r)=(1-Kr^2)(r^2-b\alpha'^2),
\end{equation}
where $b$ is an integration constant ($b\alpha'^2=E,$ in the
notation of Eqs.(2.18)-(2.19)). This equation is solved in terms of
elliptic functions and all solutions describe oscillating strings (Fig.7
shows the potential $V(r)$ for $K>0,\;K=0,\;K<0$).

For $K>0,$ when the spatial section is a hypersphere, the string either
oscillates on one hemisphere or on the full hypersphere. The energy is always
positive while the average pressure can be positive, negative or zero,
depending on the precise value of an elliptic modulus;
the equation of state is given explicitly in Table 5, in the different cases.
Interestingly enough, we find that the circular strings provide a
self-consistent solution to the Einstein equations with a selected
value of the curvature
index. Self-consistent solutions to the Einstein equations with
string sources have been found previously in the form of power law
expanding universes \cite{self}.

We have semi-classically quantized the
circular string solutions of Eq.(5.3). We used an approach
developed in field theory by Dashen et. al. \cite{das}, based
on the stationary phase approximation of
the partition function.
The result of the stationary phase integration is expressed in terms of
the function:
\begin{equation}
W(m)\equiv S_{\mbox{cl}}(T(m))+m\;T(m),
\end{equation}
where $S_{\mbox{cl}}$ is the action of the classical solution, $m$ is the
mass and
the period $T(m)$ is implicitly given by:
\begin{equation}
\frac{dS_{\mbox{cl}}}{dT}=-m.
\end{equation}
The bound state quantization condition then becomes \cite{das}:
\begin{equation}
W(m)=2\pi n,\quad n \in N_{0}
\end{equation}
for $n$ `large'. Parametric plots of $K\alpha'W$ as a function of
$K\alpha'^2m^2$ in the different cases, are shown in Fig.8. The
strings oscillating on one hemisphere give rise to a finite number $N_-$
of states with the following mass-formula Ref.\cite{all4}:
\begin{equation}
m_-^2\alpha'\approx \pi\;n,\;\;\;\;\;\;\;\;N_-\approx\frac{4}{\pi K\alpha'}.
\end{equation}
As in flat Minkowski spacetime, the scale of these string states is set by
$\alpha'.$ The strings oscillating on the full hypersphere give rise to
an infinity of more and more massive states with the asymptotic
mass-formula Ref.\cite{all4}:
\begin{equation}
m_+^2\approx K\;n^2.
\end{equation}
The masses of these states are independent of $\alpha',$ the scale is set
by the curvature index $K.$ Notice also that the level spacing grows
with $n.$ A similar result was found recently for strings in anti de
Sitter spacetime \cite{all5}.

For $K<0,$ when the spatial section is a hyperboloid, both the energy and
the average pressure of the oscillating strings are positive. The equation
of state is given in Table 5.
In this case, the strings can not provide a
self-consistent solution to the Einstein equations. After semi-classical
quantization, we find an infinity of more and more massive states. The
mass-formula is given by (see Ref.\cite{all4}):
\begin{equation}
\sqrt{-Km^2\alpha'^2}\;\log\sqrt{-Km^2\alpha'^2}\approx
-\frac{\pi}{2}K\alpha'\;n
\end{equation}
Notice that the level spacing grows faster than in
Minkowski spacetime but slower than in the closed
static Robertson-Walker spacetime.

On the other hand, the stationary strings are determined by
Eqs.(4.4):
\begin{equation}
\phi'=\frac{L}{r^2},\;\;\;\;\;\;\;\;
r'^2+V(r)=0;\;\;\;V(r)=(1-Kr^2)(\frac{L^2}{r^2}-1),
\end{equation}
where $L$ is an integration constant (Fig.9 shows the potential
$V(r)$ for $K>0,\;K=0,\;K<0$).

For $K>0,$ all the stationary string
solutions describe circular strings winding around the hypersphere.
The equation of state is of the extremely unstable string type \cite{ven}.
For $K<0,$ the stationary strings are represented by infinitely long
open configurations with an angle between the two "arms" given by:
\begin{equation}
\Delta\phi=\pi-2\arctan(\sqrt{-K}L).
\end{equation}
The energy density is positive while the pressure densities are negative.
No simple equation of state is found for these solutions.

We also computed the first and second order fluctuations
around a static string center of mass, using the string perturbation series
approach \cite{veg1}
and its covariant versions \cite{men,all1}. Up to second order,
the mass-formula for arbitrary values of the curvature index
(positive or negative) is identical to the well-known flat spacetime
mass-formula; all dependence on $K$ cancels out.
\section{Conclusion}
We first studied the string propagation in the 2+1 BH-AdS
background. We found the first and second order perturbations around the
string center of mass as well as the mass formula, and compared with the
ordinary black hole AdS spacetime and with the black string background as
well.
These results were then generalized to ordinary $D$-dimensional anti de
Sitter spacetime. The classical string motion in
anti de Sitter spacetime is stable in the sense
that it is oscillatory in time with real frequencies and the string
size and energy are bounded. Quantum mechanically, this reflects in the mass
operator, which is well defined for any value of the wave number $n,$ and
arbitrary high mass states (and therefore an infinite number of states) can
be constructed. This is to be contrasted with de Sitter spacetime, where
string instabilities develop, in the sense that the string size and energy
become unbounded for large de Sitter radius. For low mass states (the stable
regime), the mass operator in de Sitter spacetime is given by Eq.(3.1) but with
\begin{eqnarray}
\Omega_n(\lambda)_{\mbox{dS}}=
\frac{2n^2-m^2\alpha'\lambda}{\sqrt{n^2-m^2\alpha'\lambda}}.
\nonumber
\end{eqnarray}
Real mass solutions can be defined only up to some {\it maximal mass}
of the order $m^2\alpha'\approx 1/\lambda\;$ \cite{veg1}.
For $\lambda<<1,$ real mass solutions can be defined
only for $N\leq N_{\mbox{max}}\sim 0.15/\lambda$ (where $N$ is the
eigenvalue of the number operator) and therefore there exists
a {\it finite} number of states only. These features of strings in
de Sitter spacetime have been recently confirmed within a different
(semi-classical) quantization
approach based on {\it exact} circular string solutions \cite{all5}.

The presence of a cosmological constant $\Lambda$
(positive or negative) increases
considerably the number of levels of different eigenvalue of the
mass operator (there is a splitting of levels) with
respect to flat spacetime. That is, a non-zero cosmological
constant {\it decreases}
(although does not remove) the degeneracy of the string mass states,
introducing
a {\it fine structure effect}. For the low mass states the level spacing is
approximately constant (up to corrections of the order $\lambda^2$). For the
high mass states, the changes are more drastic and they depend crucially on
the sign of $\Lambda.$ A value $\Lambda<0$ causes the {\it growing}
of the level
spacing linearly with $N$ instead of being constant as in Minkowski
space. Consequently, the density of states $\rho(m)$ grows with the
exponential of $\sqrt{m}$ (instead of $m$ as in Minkowski space) discarding the
existence of a Hagedorn temperature in AdS spacetime, and the
possibility of a phase transition. In addition, another important feature of
the high mass string spectrum in AdS spacetime
is that it becomes independent of
$\alpha'.$ The string scale for the high mass states
is given by $\mid\Lambda\mid,$
instead of $1/\alpha'$ for the low mass states.

We have found new classes of
exact string solutions obtained either by the circular string
ansatz, Eq.(2.16), or by the
stationary string ansatz, Eq.(4.3),
in a variety of curved backgrounds including
Schwarzschild, de Sitter and anti de Sitter spacetimes. Many different
types of solutions have been found: oscillating circular strings,
closed stationary strings,
infinitely long stationary strings, dynamical straight strings and
multi-string solutions describing finitely or infinitely many stationary
or dynamical strings. In all the cases we have obtained the exact solutions in
terms of either elementary or elliptic functions. Furthermore, we have
analyzed the physical properties (length, energy, pressure) of the string
solutions.

Finally, we solved the equations of motion and constraints for circular strings
in static Robertson-Walker spacetimes. We computed the equations of state
and found that there exists
a self-consistent solution to the Einstein equations in the case of
positive spatial curvature. The solutions
were quantized semi-classically using the stationary phase approximation
method, and the resulting spectra were analyzed and discussed. We
have also found all
stationary string configurations in these spacetimes and we computed the
corresponding physical quantities, string length, energy and pressure.
\vskip 12pt
\hspace*{-6mm}{\bf Acknowledgements:}\\
A.L. Larsen is supported by the Danish Natural Science Research
Council under grant No. 11-1231-1SE

\newpage
\begin{centerline}
{\bf Figure Captions}
\end{centerline}
\vskip 24pt
\hspace*{-6mm}Figure 1. The potential $V(r),$ Eq.(2.21),
for a circular string in the $2+1$ BH-AdS
spacetime. In (a) we have $J^2>4E^2$ and a barrier
between the inner horizon and $r=0$, while (b) represents a case where
$J^2<4E^2$ and a string will always fall into $r=0$.
The static limit is $r_{\mbox{erg}}=l$.
\vskip 12pt
\hspace*{-6mm}Figure 2. The potential $V(r),$ Eq.(2.18), for
a circular string in the
equatorial plane of the four $3+1$ dimensional spacetimes: (a) Minkowski
(Min), (b) Schwarzschild black hole (S), (c) anti de Sitter (AdS),
and (d) Schwarzschild
anti de Sitter space (S-AdS).
\vskip 12pt
\hspace*{-6mm}Figure 3. The $(N,M)=(5,1)$ multi-string solution in anti de
Sitter spacetime. The $(N,M)$ multi-string solutions describe $N$
stationary strings with $M$ windings in the Azimuthal angle $\phi,$
in anti de Sitter spacetime.
\vskip 12pt
\hspace*{-6mm}Figure 4. The $(N,M)=(3,2)$ stationary string solution inside the
horizon of de Sitter
spacetime. Besides the circular string, this is the simplest stationary
closed string configuration in de Sitter spacetime.
\vskip 12pt
\hspace*{-6mm}Figure 5. Schematic representation of the
time evolution of the $(N,M)=(5,1)$ dynamical multi-string solution, outside
the horizon of de Sitter spacetime. Only one of the 5 strings is shown;
the others are obtained by rotating the figure by the angles
$2\pi/5,\;4\pi/5,\;6\pi/5$ and $8\pi/5.$ During the "scattering" at the
horizon, the string collapses to a point and re-expands.
\vskip 12pt
\hspace*{-6mm}Figure 6. The dynamical straight string inside the horizon of
the Schwarzschild black hole. The string infalls non-radially
towards the singularity with indefinetely growing length.
\vskip 12pt
\hspace*{-6mm}Figure 7. The potential $V(r)$ introduced in Eqs.(5.3)
for a circular string in the static Robertson-Walker spacetimes:
(a) flat $(K=0),$ (b) closed $(K>0\;\;and\;\;bK\alpha'^2\leq 1),$
(c) closed $(K>0\;\;and\;\;bK\alpha'^2>1),$
(d) hyperbolic $(K<0).$
\vskip 12pt
\hspace*{-6mm}Figure 8. Parametric plot
of $K\alpha'W$ as a function of $K\alpha'^2m^2$ in
the three cases: (a) $K>0,\;bK\alpha'^2\leq 1$ (strings oscillating on one
hemisphere, only finitely many states),
(b) $K>0,\;bK\alpha'^2>1$ (strings oscillating on the full
hypersphere, infinitely many states),
(c) $K<0$ (strings oscillating on the hyperboloid, infinitely many states).
\vskip 12pt
\hspace*{-6mm}Figure 9. The potential $V(r)$ introduced in Eqs.(5.10)
for a stationary string in the static Robertson-Walker spacetimes:
(a) flat $(K=0),$ (b) closed $(K>0\;\;and\;\;KL^2\leq 1),$
(c) hyperbolic $(K<0).$
\newpage
\begin{centerline}
{\bf Table Captions}
\end{centerline}
\vskip 24pt
\hspace*{-6mm}Table 1. String
motion described by the string perturbation series approach in
the 2+1 BH-AdS, ordinary black hole-AdS, de Sitter (dS) and
black string backgrounds.
\vskip 12pt
\hspace*{-6mm}Table 2. Circular exact string solutions in the indicated
backgrounds. $S(\tau)$ is the invariant string size.
\vskip 12pt
\hspace*{-6mm}Table 3. Characteristic
features of the quantum string mass spectrum
in anti de Sitter (AdS) and de Sitter (dS)
spacetimes.
\vskip 12pt
\hspace*{-6mm}Table 4. Short summary of the features of the string
solutions found in this section. In anti de Sitter spacetime and outside
the horizon of de Sitter and $2+1$ BH-AdS spacetimes, the
solutions describe a {\it finite} number of strings provided a condition
of the form $N\Delta\phi=2\pi M$ is fulfilled, where $\Delta\phi$ is the
angle betwen the "arms" of the string and $(N,M)$ are integers.
\vskip 12pt
\hspace*{-6mm}Table 5. Classical circular strings in the static
Robertson-Walker spacetimes. Notice that a self-consistent solution to the
Einstein equations, with the string back-reaction included, can be obtained
only for $K>0.$
\vskip 12pt
\hspace*{-6mm}Table 6. Semi-classical quantization of the circular strings
in the static Robertson-Walker spacetimes. Notice in particular the
different behaviour of the high mass spectrum of strings in the three cases.
\vskip 12pt
\hspace*{-6mm}Table 7. Stationary strings in the static
Robertson-Walker spacetimes. Notice that the pressure densities are
always negative in all three cases.

\newpage
\begin{thebibliography}{11}
\bibitem{san3}H.J. de Vega and N. S\'{a}nchez, Phys. Rev. D42 (1990) 3969.\\
              H.J. de Vega, M. Ram\'{o}n-Medrano
              and N. S\'{a}nchez, Nucl. Phys.
              B374 (1992) 405.
\bibitem{san4}H.J. de Vega and N. S\'{a}nchez, Phys. Lett. B244 (1990) 215,
              Phys. Rev. Lett. 65C (1990) 1517, IJMP A7 (1992) 3043, Nucl.
Phys.
              B317 (1989) 706 and ibid 731.\\
              D. Amati and C. Klimcik, Phys. Lett. B210 (1988) 92.\\
              M. Costa and H.J. de Vega, Ann. Phys. 211 (1991) 223 and
              ibid 235.\\
              C. Loust\'{o} and N. S\'{a}nchez, Phys. Rev. D46 (1992) 4520.
\bibitem{mic1}V.E. Zakharov and A.V. Mikhailov, JETP 47
             (1979) 1017.\\
             H. Eichenherr, in "Integrable Quantum Field Theories", ed.
             J. Hietarinta and C. Montonen (Springer, Berlin, 1982).
\bibitem{four}H.J. de Vega and N. S\'{a}nchez, Phys. Rev. D47 (1993) 3394.
\bibitem{bar}I. Bars and K. Sfetsos, Mod. Phys. Lett. A7
             (1992) 1091.
\bibitem{six}H.J. de Vega, J.R. Mittelbrunn, M.R. Medrano and N. S\'{a}nchez,
             Phys. Lett. B232 (1994) 133 and
             "The Two-Dimensional Stringy Black Hole: a new Approach and a
             Pathology", PAR-LPTHE 93/14.
\bibitem{veg1}H.J. de Vega and N. S\'{a}nchez, Phys. Lett. B197 (1987) 320.
\bibitem{san1}H.J. de Vega and N. S\'{a}nchez, Nucl. Phys. B299 (1988) 818.
\bibitem{men}P.F. Mende, in "String Quantum Gravity and Physics at the
             Planck Energy Scale". Proceedings of the Erice Workshop held in
             June 1992, ed. N. S\'{a}nchez (World Scientific, 1993).
\bibitem{fro}A.L. Larsen and V.P. Frolov, Nucl. Phys. B414 (1994) 129.
\bibitem{mic}H.J. de Vega, A.V. Mikhailov and N. S\'{a}nchez, Teor. Mat.
             Fiz. 94 (1993) 232, Mod. Phys. Lett. A29 (1994) 2745.
\bibitem{com}F. Combes, H.J. de Vega, A.V. Mikhailov and N. S\'{a}nchez,
             Phys. Rev. D50 (1994) 2754.
\bibitem{veg4}H.J. de Vega, A.L. Larsen and N. S\'{a}nchez,
             Nucl. Phys. B427 (1994) 643.
\bibitem{kri}I. Krichever, "Two-Dimensional Algebraic-Geometrical Operators
             with Self-Consistent Potentials", Landau Institute Preprint,
             March, 1994, to appear in Funct. Analisis and Appl.
\bibitem{san2}H.J. de Vega and N. S\'{a}nchez, Nucl. Phys. B309 (1988) 552
              and ibid, 577.
\bibitem{lou}C.O. Loust\'{o} and N. S\'{a}nchez, Phys. Rev. D47 (1993) 4498.
\bibitem{ini}H.J. de Vega and I.L. Egusquiza, Phys. Rev. D50 (1994) 763.
\bibitem{all}A.L. Larsen, Phys. Rev. D50 (1994) 2623.
\bibitem{axe}A.L. Larsen and M. Axenides, Phys. Lett. B318 (1993) 47.
\bibitem{self}H.J. de Vega and N. S\'{a}nchez, Phys. Rev. D50 (1994) 7202.
\bibitem{ven}N. S\'{a}nchez and G. Veneziano, Nucl. Phys. B333 (1990) 253.\\
             M. Gasperini, N. S\'{a}nchez and G. Veneziano, IJMP A6 (1991)
             3853, Nucl. Phys. B364
             (1991) 365.
\bibitem{san5}H.J. de Vega, M. Ram\'{o}n-Medrano
              and N. S\'{a}nchez, Nucl. Phys.
              B374 (1992) 425.
\bibitem{san6}H.J. de Vega and N. S\'{a}nchez, Phys. Rev. D45 (1992) 2783.\\
              M. Ram\'{o}n-Medrano and N. S\'{a}nchez, Class. Quantum Grav.
              10 (1993) 2007.
\bibitem{san7}N. S\'{a}nchez, Phys. Lett. B195 (1987) 160.
\bibitem{san8}H.J. de Vega, M. Ram\'{o}n-Medrano and N. S\'{a}nchez, Nucl.
              Phys. B351 (1991) 277.
\bibitem{all1}A.L. Larsen and N. S\'{a}nchez, Phys. Rev. D50 (1994) 7493.
\bibitem{all2}A.L. Larsen and N. S\'{a}nchez, "Mass Spectrum of Strings in
              Anti de Sitter Spacetime", DEMIRM-Paris-94048, to appear in
              Phys. Rev. D.
\bibitem{all3}A.L. Larsen and N. S\'{a}nchez, "New Classes of Exact
              Multi-String Solutions in Curved Spacetimes",
              DEMIRM-Paris-95003, to appear in Phys. Rev. D.
\bibitem{all4}A.L. Larsen and N. S\'{a}nchez, "The Effect of Spatial
              Curvature on the Classical and Quantum Strings",
              DEMIRM-Paris-95004.
\bibitem{ban1}M. Ba\~{n}ados, C. Teitelboim and J. Zanelli, Phys. Rev. Lett.
              69 (1992) 1849.
\bibitem{hor1}G.T. Horowitz and D.L. Welch, Phys. Rev. Lett. 71 (1993) 328.
\bibitem{kal}N. Kaloper, Phys. Rev. D48 (1993) 2598.
\bibitem{ban2}M. Ba\~{n}ados, M. Henneaux, C. Teitelboim and J. Zanelli,
              Phys. Rev. D48 (1993) 1506.
\bibitem{far}C. Farina, J. Gamboa and A.J. Segu\'{i}-Santonja, Class.
             Quantum Grav. 10 (1993) L193.\\
             N. Cruz, C. Mart\'{i}nez and
             L. Pe\~{n}a, Class. Quantum Grav. 11 (1994) 2731.
\bibitem{delta}A. Ach\'{u}carro and M.E. Ortiz, Phys. Rev. D48 (1993) 3600.\\
               D. Cangemi, M. Leblanc and R.B. Mann,
               Phys. Rev. D48 (1993) 3606.
\bibitem{hor2}J. Horne and G.T. Horowitz, Nucl. Phys. B368 (1992) 444.
\bibitem{bus}T. Buscher, Phys. Lett. B201 (1988) 466, B194 (1987) 59.
\bibitem{vil4}A. Vilenkin, Phys. Rev. D24 (1981) 2082.
\bibitem{energy}H.J. de Vega and N. S\'anchez,
              Int. J. Mod. Phys. A7 (1992) 3043.
\bibitem{fro1}V.P. Frolov, V.D. Skarzhinsky, A.I. Zelnikov and O. Heinrich,
              Phys. Lett. B224 (1989) 255.
\bibitem{das}R. Dashen, B. Hasslacher and A. Neveu, Phys. Rev. D11 (1975) 3424.
\bibitem{all5}H.J. de Vega, A.L. Larsen and N. S\'{a}nchez, "Semi-Classical
              Quantization of Circular Strings in de Sitter and
              anti de Sitter Spacetimes", DEMIRM-Paris-94049, submitted to
              Phys. Rev. D.
\bibitem{all6}A.L. Larsen, "Stable and Unstable Circular Strings in
              Inflationary Universes", NORDITA-94/14-P, to appear in
              Phys. Rev. D.
\bibitem{all7}A.L. Larsen, Nucl. Phys. B412 (1994) 372.
\end{thebibliography}
\end{document}